# A Silicon Beam Tracker

J. H. Han, H. S. Ahn, J. B. Bae, H. J. Hyun, S. W. Jung, D. H. Kah, C. H. Kim, H. J. Kim, K. C. Kim, M. H. Lee, L. Lutz, A. Malinin, H. Park, S. Ryu, E. S. Seo, P. Walpole, J. Wu, J. H. Yoo, Y. S. Yoon, S. Y. Zinn

*Abstract*—When testing and calibrating particle detectors in a test beam, accurate tracking information independent of the detector being tested is extremely useful during the offline analysis of the data. A general-purpose Silicon Beam Tracker (SBT) was constructed with an active area of 32.0 × 32.0 mm$^2$ to provide this capability for the beam calibration of the Cosmic Ray Energetics And Mass (CREAM) calorimeter. The tracker consists of two modules, each comprised of two orthogonal layers of 380 μm thick silicon strip sensors. In one module each layer is a 64-channel AC-coupled single-sided silicon strip detector (SSD) with a 0.5 mm pitch. In the other, each layer is a 32-channel DC-coupled single-sided SSD with a 1.0 mm pitch. The signals from the 4 layers are read out using modified CREAM hodoscope front-end electronics with a USB 2.0 interface board to a Linux DAQ PC. In this paper, we present the construction of the SBT, along with its performance in radioactive source tests and in a CERN beam test in October 2006.

## I. Introduction

Cosmic Ray Energetics And Mass (CREAM) is a balloon-borne experiment to study cosmic-ray particles from protons to iron nuclei at energies up to ~ $10^{15}$ eV [1, 2]. A sampling tungsten/scintillating fiber calorimeter in the experiment measures energies of cosmic-ray nuclei by looking at the scintillation signal from showers of secondary charged particles in the calorimeter volume when particles interact inelastically [3]. The calorimeter was exposed to electron beams of known energy, i.e. 150 GeV at CERN's SPS facility for calibration purposes. Independent tracking information of beam particles on an event by event basis is very useful for this test since the calorimeter active layers are segmented into 1 cm wide scintillating fiber ribbons for optimal tracking. A Silicon Beam Tracker (SBT) made of silicon strip sensors [4, 5] and hodoscope electronics [6] was constructed to provide this independent beam tracking. A USB 2.0 interface controller board was designed to control and power the front-end electronics and to provide data interface to the DAQ PC. The SBT system, including the USB 2.0 controller can be used in any beam test as a general-purpose, portable beam tracker. This paper details the design and construction of the SBT system, and discusses its performance in radioactive source tests during construction, as well as in the beam test.

## II. Design and construction of the Silicon Beam Tracker

### A. Sensor

Two types of single-sided silicon sensors of 32 × 32 mm$^2$ active areas were used: a DC-type sensor with 32 strips of 1 mm pitch, and an AC-type sensor with 64 strips of 0.5 mm pitch. Table 1 summarizes the specifications of these sensors.

TABLE I
SPECIFICATIONS OF SBT SENSORS

|  | DC-coupled | AC-coupled |
|---|---|---|
| Sensor Area (μm$^2$) | 35000 × 35000 | 35000 × 35000 |
| Effective Area (μm$^2$) | 31970 × 31970 | 31970 × 31970 |
| Thickness (μm) | 380 | 380 |
| Number of strips | 32 | 64 |
| Strip pitch (μm) | 1000 | 500 |
| Strip width (μm) | 400 | 200 |
| Readout width (μm) | 420 | 220 |

### B. Detector board

A detector board was designed to mount a silicon sensor with wire bonding between sensor strips and readout circuit traces. Two types of detector boards were produced; one is for each type of sensors. The DC detector board includes coupling capacitors and resistors, while the AC-type sensor incorporates these on-chips, making for a simpler readout circuit (Fig. 1). Each sensor was mounted on its detector board using conductive epoxy (XCA3556) applied only at the two ground sides of the sensors, along the strip direction, while 25 μm gold plated aluminum wire was used to bond each strip except the two at the edges. Board traces are also gold plated for compatibility. The two unconnected channels, lacking genuine signals, are used to monitor coherent behavior for noise reduction in offline data analysis. A glob top of DP100 epoxy was applied at each wire bonding site for mechanical protection. Mechanical support and light-shielding are provided

This work was supported by NASA, by the KOSEF under a BAERI program and by the International Cooperation Program of the KICOS.

J. H. Han, H. S. Ahn, C. H. Kim, K. C. Kim, M. H. Lee, L. Lutz, A. Malinin, E. S. Seo, P. Walpole, J. Wu, J. H. Yoo, Y. S. Yoon, and S. Y. Zinn are with the Institute for Physical Science and Technology, University of Maryland, College Park. MD 20742 USA (e-mail: mhlee@umd.edu).

J. B. Bae, H. J. Hyun, S. W. Jung, D. H. Kah, H. J. Kim, H. Park, and S. Ryu are with the Department of Physics, Kyungpook National University, Daegu, 702-701, Republic of Korea

E. S. Seo, and Y. S. Yoon are with the Department of Physics and the Institute for Physical Science and Technology, University of Maryland, College Park. MD 20742 USA by an aluminum frame on one side of the board, attached to a Delrin plastic frame on the other side. The frames both have reduced thickness windows (0.5 mm thick) over the sensor. Five detector boards of each type were produced.

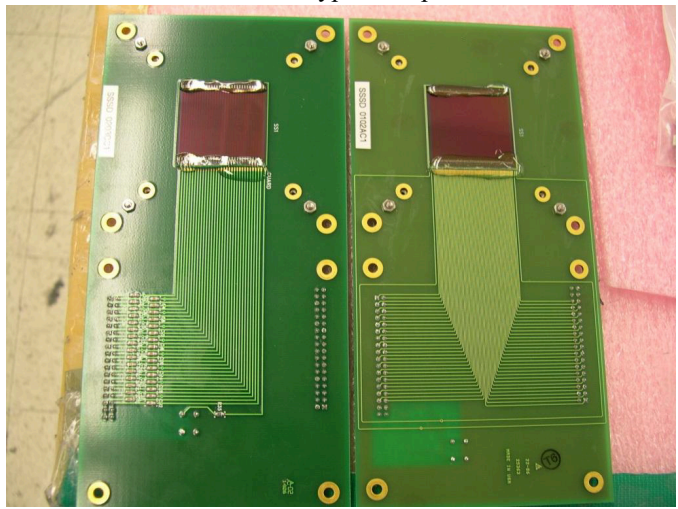

Fig. 1. Two types of detector boards with sensors mounted, side by side before assembly. The DC-type (left) has coupling capacitors and resistors on the board; the AC-type (right) has none.

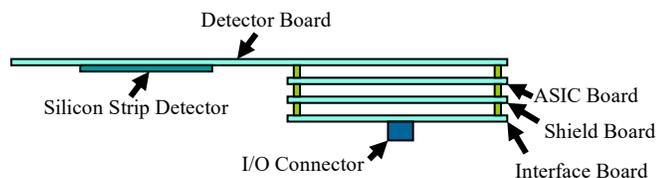

Fig. 2. Side-view schematic of an SBT detector board assembled with a set of modified hodoscope front end electronics boards.

A complete SBT unit (Fig. 3) is comprised of two detector boards, each with its front-end readout circuitry. The two detectors measure orthogonal directions, providing the two coordinates needed for position measurement at a beam test.

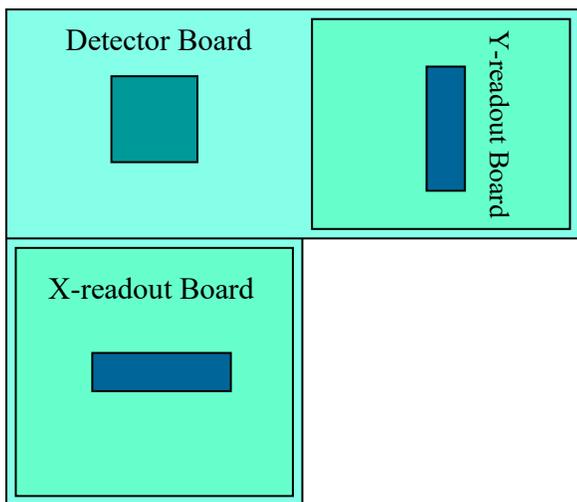

Fig. 3. Beam-axis schematic view of an assembled SBT unit, comprised of two detector board assemblies for two-coordinate measurements.

## III. USB CONTROLLER BOARD

A USB 2.0 interface controller board was designed to make the SBT system as flexible and portable as possible. As shown in Fig. 4, a USB controller board, with 256 channels, can connect up to 4 detector boards to the DAQ PC in parallel. The controller board suppresses or sparsifies those data below threshold values set by commands from the DAQ PC, and provides bias voltages to the four detector boards. An adjustable peak hold signal is generated for the front-end electronics by receiving direct trigger inputs in NIM or LVDS standards. The controller receives a test pulse signal and distributes it to the front-end electronics of the 4 detector boards to provide charge calibration. An event number from a master trigger board is recorded in each event data record to synchronize with other instruments being tested. This general-purpose controller board can be used in most circumstances with a USB 2 port in the DAQ PC.

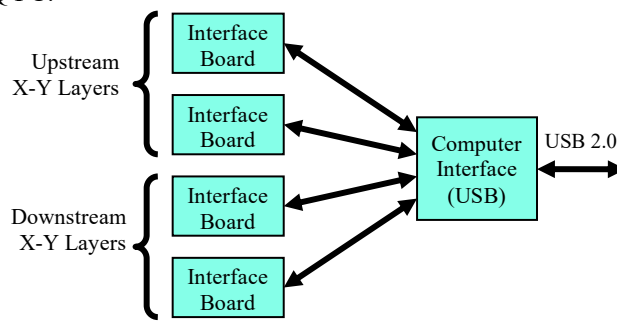

Fig. 4. Diagram of USB controller board

## IV. RADIOACTIVE SOURCE TEST

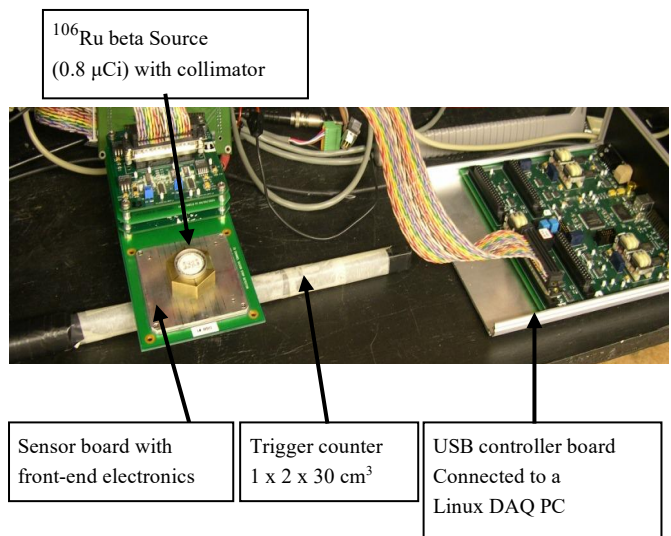

Fig. 5. A $^{106}$Ru source test setup of a detector board assembly with a USB controller board

During the construction of the SBT, various tests were carried out using a radioactive source to tune bias voltages and verify performance.

### A. Test setup

Figure 5 shows a test setup for a detector board assembly with a collimated $^{106}$Ru beta source (Maximum E = 3.541 MeV). A $1 \times 2 \times 30$ cm$^3$ plastic scintillation counter was used as a trigger counter to select minimum ionizing electrons in the silicon sensor by requiring relatively high threshold values. Since electrons lose at least 2 MeV/cm in plastic, the 2 cm thick scintillator absorbs the full energy of these beta electrons, so such a threshold correctly identifies high energy electrons.

### B. Bias scan

The signals from a detector strip are plotted using the above test setup (Fig. 6), after coherent noise subtraction using the unconnected channels in various bias voltages from 10 to 90 V.

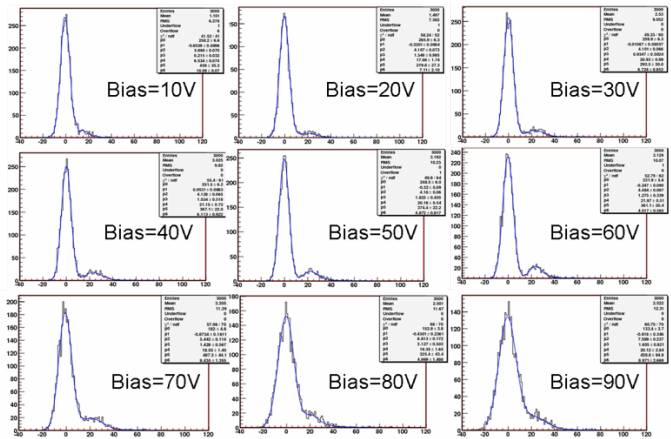

Fig. 6. Pedestal and signal distributions at different bias voltage from a strip.

### C. Optimal bias value

A detector board assembly was tested with different bias voltages to find the optimal voltage for operation. The most probable values of the signal distributions were plotted as a function of bias voltage (Fig. 7). The silicon sensor is shown to be fully depleted at 30 V as expected from the higher resistance of the sensor. As the voltage exceeds 70 V, noise levels start increasing. All 10 detector board assemblies were tested with the source to select the four with the highest S/N for the SBT. After the selection tests, the bias voltage was set at 50 V as the optimal value for the 4 sensors.

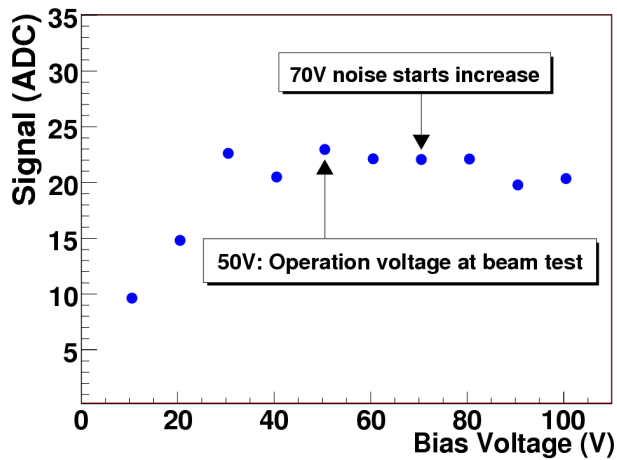

Fig. 7. Most probable values of pedestal-subtracted signals at various bias voltages

### D. Signal to Noise ratio

Large statistics data from a detector strip are plotted in Fig. 8. The measured signal to noise ratio (S/N) is ~5.4.

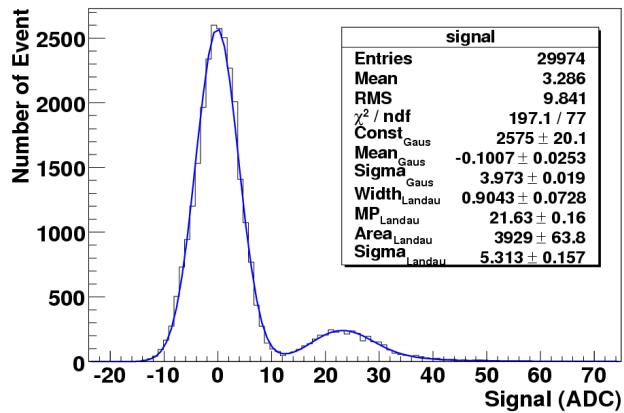

Fig. 8. A distribution of signals from a strip excited by a beta source. The noise level is ~4 ADC channels while the signal is ~22 ADC channels, providing S/N of ~5.4

### E. Position scan

In the test setup, the source is moved about 6 mm and saw the difference in RMS of the ADC values for each strip before and after the movement as shown in Fig. 9. With the source profile is about 5.5 mm wide, a clear separation of two profiles are seen. It shows that this detector can be used as a simple and fast charged beam profile monitor in a beamline by just looking at the RMS of the on-line data.

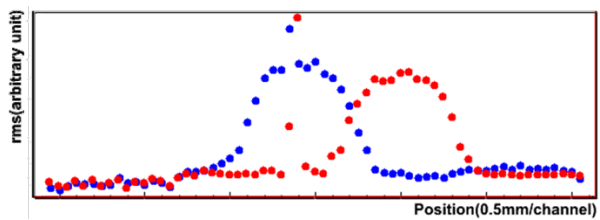

Fig. 9. Source profiles from two different source positions, apart by ~ 6 mm, are well separated by just looking at the rms of ADC values

## V. PERFORMANCE AT THE BEAM TEST

The two SBT units were installed in a CERN SPS beamline and used as a beam tracker during a CREAM calorimeter module calibration run in October of 2006. The two units were positioned 60 cm apart from each other to provide a sufficient lever arm for tracking, with the nearer about 2.5 m upstream from the calorimeter. Good hits were selected by requiring that the strip signals are at least 3.5 σ above the pedestal mean value. Multi-track events were removed by requiring ≤ 2 hits/sensor. The hits from two sensors in the same orientation were fitted with a straight line. Since the sensors were mounted perpendicular to the beam, only track candidates with a vertical direction were retained as final tracks. The SBT tracking provided critical independent position information for calorimeter calibration. Figure 8 shows an example of beam profiles obtained from two sensors in the y-direction.

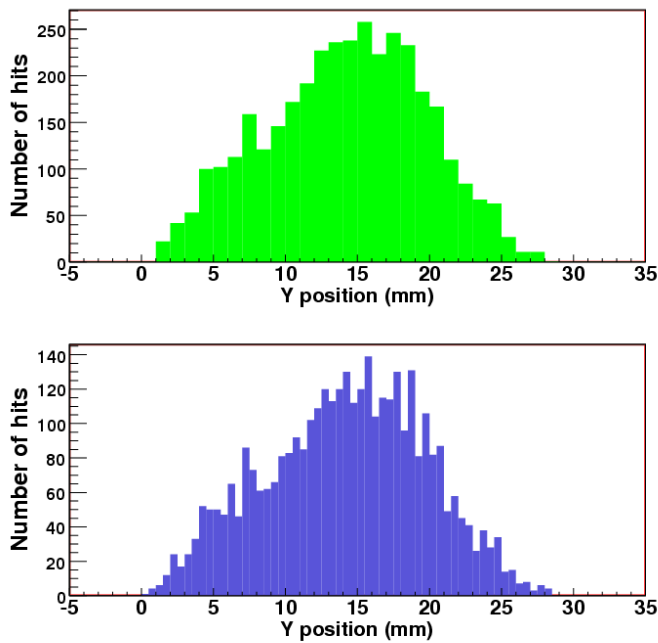

Fig. 10. 150 GeV electron beam profile measured using the 1 mm pitch sensor (top) and the 0.5 mm pitch sensor (bottom)

## VI. CONCLUSION

A general-purpose portable Silicon Beam Tracker system was designed, constructed, and tested with a radioactive source and particle beams. Using modified CREAM hodoscope front-end electronics and a newly designed USB 2.0 interface controller, a signal to noise ratio (S/N) of ~5.4 was achieved for minimum ionizing particles. In a CREAM beam test at CERN, 4 SBT assemblies were used to provide two sets of x and y measurements. The SBT data show a good beam profile.

## ACKNOWLEDGMENT

We thank CERN for their excellent test beam facility and services.